\documentstyle[aps]{revtex}
\begin{document}
 \draft
\def\lap{\Delta}                        

\twocolumn[\hsize\textwidth\columnwidth\hsize\csname
@twocolumnfalse\endcsname
\preprint{SU-ITP-99-\\ hep-th/ \\  1999}
\date{April 1999}
\title{Anomaly induced effective actions and Hawking radiation }
\author{Roberto Balbinot
 }
\address{Dipartimento di Fisica dell'Universit\`a di Bologna and INFN 
sezione di Bologna, Via Irnerio 46, 40126 Bologna, Italy}
\author{Alessandro Fabbri}
\address{
Department of Physics, Stanford University, Stanford,
CA 94305-4060, USA  }
\author{Ilya Shapiro}
\address{ Departamento de Fisica, ICE, Universidade Federal de Juiz de Fora, 
Juiz de Fora, MG, Brazil\\
Tomsk Pedagogical University, Tomsk, Russia}
\maketitle
\begin{abstract}
The quantum stress tensor in the Unruh 
state for a conformal scalar 
propagating in a 4D Schwarzschild black hole spacetime is reconstructed
in its leading behaviour at infinity and near the horizon by means of an 
effective action derived by functionally integrating the trace anomaly.
 \end{abstract}

\pacs{PACS: 04.70.Dy, 04.62+v 
\hskip 2.0cm SU-ITP-99-16
}
\vskip2pc]

In the mid-seventies Hawking \cite{Hawking} showed that black holes
are quantum mechanically unstable: they decay by the emission of thermal 
radiation at a temperature inversely proportional to their mass, i.e. 
$T_H=(8\pi M)^{-1}$ in units where $\hbar=c=G=k_B=1$. This is one of 
the most astonishing discoveries of theoretical physics in the second 
half of the century. 
Nowadays black hole radiation and its thermodynamical implications, 
most notably Bekenstein-Hawking area-entropy formula \cite{bekenstein},
are among the consistency tests any candidate of quantum gravity 
theory has to successfully pass in order to be seriously considered. 
Notwithstanding decades of intensive studies, the evolution and fate of 
an evaporating black hole (EBH) are still unknown. In the opinion
of many people the final answer to this issue has to wait until a 
complete and self consistent quantum gravity theory has been found. 
String theory appears as the most promising candidate to achieve this 
goal and many efforts have been devoted to show their compatibility with 
black hole radiation. However, one is still far away to understand, 
within string theory, how do black holes evaporate. \\ \noindent
{}A more traditional field theoretical approach to the 
evolution of black holes driven by the quantum fluctuations of the matter 
fields relies on the effective action $S_{eff}(g_{\alpha\beta})$. 
This is the so called backreaction, which in mathematical 
terms is governed by the semiclassical Einstein equations
\begin{equation}
\label{1}
G_{\mu\nu}(g_{\alpha\beta})=8\pi\langle 
T_{\mu\nu}(g_{\alpha\beta})\rangle \ ,
\end{equation}
where $G_{\mu\nu}$ is the Einstein tensor for the metric 
$g_{\alpha\beta}$ and 
\begin{equation}
\label{2}
\langle T_{\mu\nu}(g_{\alpha\beta})\rangle = \frac{2}{\sqrt{-g}}
\frac{\delta S_{eff}(g_{\alpha\beta})}{\delta g^{\mu\nu}}
\end{equation}
is the renormalized expectation value of the stress energy tensor 
operator for the matter fields propagating on $g_{\alpha\beta}$.
A quantum state and boundary conditions appropriate to black hole
evaporation have to be supplied to eq. (\ref{1}). \\ \noindent
The framework is quantum field theory in curved space-time \cite{duff},
a semiclassical approach in which only the matter fields are quantized,
whereas gravity is still described classically according to Einstein's 
General Relativity. One expects this approximation to be consistent until 
the size of the EBH becomes comparable to the Planck length ($10^{-33}$cm).
At this point one has to move to a genuine quantum gravity theory 
which unfortunately is still lacking. 
Even within semiclassical gravity, however, the evolution of an EBH is 
hard to follow simply because the relevant effective action 
$S_{eff}(g_{\alpha\beta})$ is not explicitly known. The only informations 
available for black hole evaporation come from analytical estimates of
$\langle T_{\mu\nu}\rangle$ for matter fields propagating in a fixed 
static Schwarzschild black hole geometry of a given mass $M$.
Selecting a mode basis suitable for black hole evaporation (Unruh modes 
\cite{unruh} ) the matter fields are expanded in that basis, 
canonically quantized and then $\langle T_{\mu\nu}\rangle$ is directly 
calculated by modes sum and point splitting regularization of the 
divergences. \\
Note that the Schwarzschild spacetime
\begin{equation}
\label{3}
ds^2=-fdt^2 + f^{-1}dr^2 +r^2(d\theta^2 +\sin^2\theta
d\varphi^2), 
\end{equation}
where $f=1-2M/r$,
does not satisfy the semiclassical Einstein equations (\ref{1}) being the 
l.h.s. identically vanishing unlike the r.h.s. . However one can still 
regard a Schwarzschild black hole as a sort of zero order (in the hole 
luminosity) approximation to a real EBH. \\
The mode basis relevant for quantization are chosen in the following  
way: \\ \noindent
i) {\it{in}} modes are positive frequency with respect to Minkowsky time t;
\\ \noindent
ii) {\it{out}} modes are positive frequency with respect to Kruskal 
$U=-4Me^{-u/4M}$,
the affine parameter along the past horizon. \\
The quantum state so defined is called the Unruh state. 
By condition i) this state reduces to the usual Minkowski {\it in} vacuum
asymptotically in the past (i.e. no incoming radiation). The condition 
ii) mimics the modes coming out from a collapsing star as its surface 
approaches the event horizon, as shown in Hawking's original 
analysis \cite{Hawking}. \\
$\langle T_{\mu\nu}\rangle$ in the Unruh state
has
 the following leading behaviour at infinity \cite{candelas} (only the
nonzero components are shown)
\begin{equation}
\label{4}
\langle  T_{a}^{\ b}   \rangle \to  \frac{L}{4\pi r^2}
\pmatrix{ -1 & -1 
\cr 1 & 1 \cr}\ ,
\end{equation}
$a,b=r,t$,
corresponding to an outgoing flux of (approximately) blackbody radiation 
at the Hawking temperature $T_H$. $L$ is the luminosity of the black hole
and is proportional to $M^{-2}$ (for a scalar field geometric optics yields 
$L=\frac{2.197\ 10^{-4}}{\pi M^2}$ \cite{dewitt} ) .
 On the future event horizon $\langle 
T_{\mu\nu}\rangle$ is regular in a free falling frame as a consequence 
of ii) and one finds \cite{candelas} that 
$\langle T_{\theta}^{\ \theta}\rangle $ 
is finite and 
\begin{equation}
\label{5}
\langle  T_{a}^{\ b}   \rangle \sim \frac{L}{4\pi (2M)^2 }
  \pmatrix{ 1/f & -1 \\
\cr 1/f^2 & -1/f \cr} \ ,
\end{equation}
describing an influx down the hole of negative energy radiation 
which compensates the flux escaping at infinity.
From these results one expects black holes to evaporate at a rate of 
order $M^{-2}$. The evolution is then modelled as a sequence of 
Schwarzschild black holes with the mass parameter $M$ decreasing along 
the sequence at the above rate. This should hold at least to zero order. \\
To go beyond this naive scheme one should directly attack the 
semiclassical Einstein equations: find $\langle 
T_{\mu\nu}(g_{\alpha\beta}) \rangle$ for a sufficiently general (i.e. 
time dependent) EBH geometry $g_{\alpha\beta}$ and solve eqs. (\ref{1})
for the geometry. This for the moment remains a dream since, as said 
before, $S_{eff}(g_{\alpha\beta})$ and hence $\langle 
T_{\mu\nu}(g_{\alpha\beta})\rangle$ are not known. \\ 
Significant simplifications occur when the matter fields one is 
considering are conformal invariant, since then at least part of 
$S_{eff}(g_{\alpha\beta})$ can be reconstructed from the trace anomaly 
\cite{duff}. We shall call this part ``anomaly induced effective action'',
i.e. $S_{an}^{eff}$. \\
 At the classical level conformal invariance of the 
matter fields action implies vanishing trace of the corresponding energy 
momentum tensor. At the quantum level, on the other hand, the 
renormalization procedure induces a nonvanishing expectation value of the 
trace which does not depend on the quantum state in which the expectation 
value is taken. This trace anomaly is expressed completely in terms of 
geometrical objects \cite{duff}
\begin{equation}
\label{6}
\langle T^{\alpha}_{\alpha}\rangle \equiv \langle T \rangle =-
\frac{1}{(4\pi)^2}\left( aC^2 + bE +c\Box R \right) \ .
\end{equation}
In our notation $C^2\equiv 
C_{\alpha\beta\gamma\delta}C^{\alpha\beta\gamma\delta}$ is the square of 
the Weyl tensor and $E$ is an integrand of the Gauss-Bonnet topological 
term 
$E\equiv 
R_{\mu\nu\alpha\beta}R^{\mu\nu\alpha\beta}-4R_{\alpha\beta}R^{\alpha\beta}
+R^2\,$.
We remark that the origin of the trace anomaly is the renormalization
of the action of vacuum in a theory of conformal invariant 
matter fields, that is why in (\ref{6}) the 
$R^2$-term does not show up.
Finally, the numerical coefficients $a,b,c$ depend on the matter 
species considered \cite{duff}.  
The anomaly induced effective action is related to the trace anomaly by
functional integration of  
\begin{equation}
\label{8}
\frac{2}{\sqrt{-g}}
g_{\mu\nu}\frac{\delta S^{eff}_{an}}{\delta g_{\mu\nu}}=
\langle T\rangle \ .
\end{equation}
This operation allows $S^{eff}_{an}$ to be determined up to a 
Weyl invariant functional. 
\\
The basic question we would like to address in this paper is whether 
$S^{eff}_{an}$ 
by itself is sufficiently accurate to reproduce the basic 
properties of black hole evaporation and can therefore be used in the 
semiclassical Einstein equations (\ref{1}) to get some insight in the 
backreaction problem. 
To answer this question we shall explicitly test $S^{eff}_{an}$ in a 
specific example where results can be obtained in an independent way, 
namely a massless scalar field in the Unruh state propagating on a 
Schwarzschild black hole geometry. For this system we already know from 
our previous discussion the expected leading behaviour of $\langle 
T_{\mu\nu}\rangle$ at infinity and near the horizon (see eqs. (\ref{4}), 
(\ref{5}) ). \\
We shall now proceed to show that, with appropriate boundary conditions, 
$S^{eff}_{an}$ does indeed lead to a flux of radiation at infinity 
emitted by the Schwarzschild black hole in agreement with eqs. (\ref{4}),
(\ref{5}). \\
We shall work with the following local form of $S^{eff}_{an}$
\cite{jash}, \cite{reigert}, \cite{bafash} 
\begin{equation}
\label{14}
S^{eff}_{an}=-\frac{c+\frac{2}{3}b}{12(4\pi)^2}\int d^4x \sqrt{-g}R^2 +
\end{equation}
$$
\int d^4 x\sqrt{-g} \left[ \frac{1}{2}\,\phi\Delta_4\phi 
+ \phi\left( 
\,k_1 C^2 + k_2 (E-\frac{2}{3}\Box R) \,\right) \right] 
$$
$$
 +\int d^4x\sqrt{-g}\left(-\frac{1}{2}\,\psi\Delta_4\psi
+l_1 C^2 \psi \right), 
$$
where $k_1\equiv -\frac{a}{8\pi\sqrt{-b}}$ and 
$k_2\equiv \frac{\sqrt{-b}}{8\pi}$. 
We are considering the introduction 
of the auxiliary fields as a purely classical transformation 
which doesn't modify the values of $a,b,c$ in (\ref{6}).
$\Delta_4$ is the fourth order conformal operator \cite{reigert}
\begin{equation}
\label{10}
\Delta_4=\Box^2 - 2R^{\mu\nu}\nabla_{\mu}\nabla_{\nu}+
\frac{2}{3}R\Box -\frac{1}{3}(\nabla^{\mu}R)\nabla_{\mu} 
\end{equation}
and $l_1$ is an arbitrary parameter not determined by the theory. After 
elimination of the 
auxiliary fields $\phi$ and $\psi$ this expression 
reduces to the well known nonlocal form given by Reigert 
\cite{reigert} only if $l_1=\frac{a}{8\pi\sqrt{-b}}$. 
For other values of $l_1$ this no longer happens.
The difference, however, is a conformal invariant functional 
which, as said,
cannot be determined from the trace anomaly alone. \\
From eq. (\ref{14}) the equations of motion for the auxiliary 
fields are
\begin{equation}
\label{15}
\frac{1}{\sqrt{-g}}\frac{\delta S^{eff}_{an}}{\delta\phi}=
\Delta_4\phi + k_1 C^2 + k_2(E-\frac{2}{3}\Box R)=0\ ,
\end{equation}
\begin{equation}
\label{16}
\frac{1}{\sqrt{-g}}\frac{\delta S^{eff}_{an}}{\delta\psi}=
=-\Delta_4\psi +l_1 C^2 =0\ ,
\end{equation}
Introducing the traceless tensor $K_{\mu\nu}$ as
\begin{equation}
\label{17}
K_{\mu\nu}(\phi) = \frac{1}{\sqrt{-g}}\,\frac{\delta }{\delta g^{\mu\nu}}
\int d^4x\sqrt{-g}\,\left\{ \phi\Delta_4\phi \right\} 
\end{equation}
we can write 
\begin{equation}
\label{18}
\frac{2}{\sqrt{-g}}\frac{\delta S^{eff}_{an}}{\delta g^{\mu\nu}}
\equiv \langle T_{\mu\nu}\rangle = K_{\mu\nu}(\phi)-K_{\mu\nu}(\psi) 
\end{equation}
$$
-  8\nabla^{\lambda}\nabla^{\tau}Z R_{\mu\lambda\nu\tau} + 
\,g_{\mu\nu}\, Z 
R^2_{\rho\sigma\alpha\beta} 
- 4Z\,R_{\mu\rho\lambda\tau}\,{R_{\nu }}^{\rho\lambda\tau} 
$$
$$
 -\frac{4k_2}{3}\,\left[
(\nabla_\mu\nabla_\nu {\Box} \phi)
- g_{\mu\nu} ({\Box}^2 \phi) \right] + ... ,$$
where $Z\equiv (k_1+k_2)\phi +l_1\psi$
and the dots indicate terms containing either the Ricci 
tensor $R_{\mu\nu}$ or the Ricci scalar $R$. Since for our subsequent 
analysis these terms vanish identically, they are not written in detail. \\
The procedure we shall adopt is to solve the equations of motion 
(\ref{15}) and (\ref{16}) for the auxiliary fields in the background 
Schwarzschild geometry, insert then these solutions for $\phi$ and $\psi$ 
in $\langle T_{\mu\nu}\rangle$ of eq. (\ref{18}) and compare the results 
with the expected expressions eqs. (\ref{4}), (\ref{5}). \\
The problem we immediately have to face in trying to follow the above 
scheme is how to define in our framework the Unruh state, since the 
trace anomaly and hence $S^{eff}_{an}$ do not make any reference to a 
particular quantum state. Note, however, that the solution of the 
auxiliary field equation (\ref{15}) (and similarly for eq. (\ref{16}) )
is determined up to a solution of the homogeneous equation
$\Delta_4\phi=0$.
It is through this 
solution that the state dependence will be encoded. \\
The boundary conditions that characterize the Unruh vacuum which
follow from its definition (see i) and ii)) are: \\ \noindent
a) no incoming radiation from infinity; \\ \noindent
b) $\langle T_{\mu\nu}\rangle$ should be regular on the future event horizon
(in a free falling frame) . \\
Furthemore in the Unruh state $\partial_t \langle T_{\mu\nu}\rangle =0$.\\ 
The homogeneous solutions of the auxiliary fields equation of motion have 
to implement these boundary conditions in our system if we want to 
correctly describe black hole evaporation. \\
The solution for $\phi$ can be given in the following general form
$\phi(r,t)=dt + w(r)$, where
\begin{equation}
\label{21} 
\frac{dw}{dr}= \frac{B}{3} r +
\frac{2}{3}MB-\frac{A}{6}-\frac{\alpha}{72M}
\end{equation}
$$
+ 
\left( \frac{4}{3}BM^2 + \frac{C}{2M}
-AM
 -\frac{\alpha}{24}\right) \frac{1}{r-2M}
-\frac{C}{2M}\frac{1}{r} 
$$
$$
+\ln r \left[ -\frac{\alpha}{36}\frac{2M}{r(r-2M)}
-\left( \frac{A}{2M}-\frac{\alpha}{48M^2}\right)\frac{r^2}{3(r-2M)}
 \right]
$$
$$
+\ln(r-2M) \left[ \left( \frac{A}{2M}-\frac{\alpha}{48M^2}\right)
\frac{r^3-(2M)^3}{3r(r-2M)}
\right] \ 
$$
and we have defined $\alpha\equiv -48(k_1+k_2)$.
$A,B,C,d$ are constants that specify the homogeneous solution. The choice 
of a linear $t$ dependence appearing in eq. (\ref{21}) is the following. 
In the Unruh state $\langle T_{rt}\rangle \neq 0$ and this requires our 
field $\phi$ to have a time dependence otherwise $\langle T_{rt}\rangle$ 
would vanish identically. However any time dependence different from the 
linear one would imply an explicit time dependence of $\langle T_{\mu\nu} 
\rangle$ , which contradicts $\partial_t\langle T_{\mu\nu}\rangle=0$.
Any $\theta,\varphi$ dependence is forbidden by spherical symmetry.\\ 
One can express the solution for the other auxiliary field $\psi$ in a 
similar form with the obvious replacements $\alpha\to \beta \equiv 48l_1$, 
$(A,B,C,d)\to 
(A',B',C',d')$. Substituting the solutions for the auxiliary fields 
$\phi$ and $\psi$ in eq. (\ref{18}) one obtains the stress tensor 
$\langle T_{\mu\nu}\rangle$. We symbolically write 
\begin{equation}
\label{23}
\langle T_{\mu\nu}\rangle = \langle T_{\mu\nu} (\phi)\rangle +
\langle T_{\mu\nu}(\psi)\rangle
\end{equation}
dividing the contribution of each individual auxiliary field to the 
stress tensor. The boundary conditions a) and b) will be imposed on 
$\langle T_{\mu\nu}(\phi)\rangle$ and $\langle T_{\mu\nu}(\psi)\rangle$ 
separately and the physical reason for this will become clear at the end 
of our analysis. \\
Being the calculation of $\langle T_{\mu\nu}\rangle$ 
rather lengthy and boring we shall report, here, only the basic results. 
A detailed analysis and discussion will be reported in a forthcoming 
publication \cite{bafash}. \\
The $(r,t)$ component is the most simple to 
write and reads
\begin{equation}
\label{24}
\langle T_t^{\ r}(\phi)\rangle =-\frac{dA}{r^2}\ ,
\end{equation}  
as expected from the conservation equations 
$\nabla^{\mu}\langle T_{\mu}^{\ \nu}\rangle=0$ 
in the Schwarzschild spacetime under the 
hypothesis $\partial_t \langle T_{\mu\nu}\rangle=0$ \cite{chrifu}.\\
Examining the behaviour on the horizon $r=2M$ we have 
\begin{equation}
\label{25}
\partial_r\phi \sim \frac{E}{r-2M} +
\left(A-\frac{\alpha}{24M}\right)\ln(r-2M) + reg. \ ,
\end{equation}
where
$E=-\frac{\alpha}{24}+\frac{4}{3}BM^2 +\frac{C}{2M}
-AM -\frac{2}{3}AM \ln2M \ .$\\

All logarithmic divergences in $\langle T_{\mu\nu}\rangle$ are 
eliminated by choosing 
$A=\alpha/24M$ and the leading divergence on the horizon 
then becomes 
\begin{equation}
\label{27}
\langle T_{\mu}^{\ \nu}(\phi)\rangle \sim \frac{(E^2-4d^2M^2)}{32M^4f^2}
diag(-1, 1/3, 1/3, 1/3)\ ,
\end{equation}
where as usual 
$f\equiv 1-2M/r$. This divergence vanishes if we choose $E=2dM$ and 
we find 
\begin{equation}
\label{28}
\langle  T_{a}^{\ b}(\phi)   \rangle \sim  \frac{1}{(2M)^2}
\pmatrix{ dA/f & -dA 
\cr dA/f^2 & -dA/f \cr}
\end{equation}
and $\langle T_{\theta}^{\ \theta}\rangle$ finite, 
 which yields $\langle T_{\mu}^{\ \nu}\rangle$
 regular on the future horizon as required by 
condition b). Had we chosen $E=-2dM$, which still makes eq. (\ref{27}) 
vanishing, the resulting $\langle T_{\mu}^{\ \nu}\rangle$ would be regular on 
the past horizon (and not on the future). \\
Examining the behaviour at 
infinity we find that imposing $B=0$ the leading term as $r\to\infty$ reads 
\begin{equation}
\label{31}
\langle T_{a}^{\ b}(\phi)\rangle \to \frac{1}{r^2}
\pmatrix{ -A^2/2 & -dA 
\cr dA & A^2/2 \cr}
\end{equation}
and $\langle T_{\theta}^{\ \theta}\rangle =0$ at this order.
Requiring no incoming radiation forces us to set $d=A/2$. \\
Repeating the 
steps for the other auxiliary field $\psi$ we eventually arrive at the 
final results
\begin{equation}
\label{32}
\langle T_{a}^{\ b}\rangle \to  
\frac{\alpha^2 - \beta^2}{2r^2(24M)^2} \pmatrix{ -1 & -1 
\cr 1 & 1 \cr} \ , \ r\to \infty ,
\end{equation}
\begin{equation}
\label{33}
\langle  T_{a}^{\ b}   \rangle \sim \frac{\alpha^2 - \beta^2}{2(48M^2)^2}
  \pmatrix{ 1/f & -1 \\
\cr 1/f^2 & -1/f \cr} \ ,\ r\to 2M .
\end{equation}
It is remarkable that 
these expressions are exactly the required form of eqs. (\ref{4}), 
(\ref{5}) if we set $\frac{L}{4\pi}=\frac{\alpha^2 - \beta^2}{2(24M)^2} $.\\ 
Before proceeding to a numerical 
comparison, it is rather illuminating to examine the analytic structure 
of the auxiliary fields once the arbitrary constants $(A,B,C,d)$ and 
$(A',B',C',d')$ are fixed according to our Unruh state conditions a) and 
b). As $r\to 2M$ we find that the condition $E=2dM$
makes $\phi$ linear in $v$, i.e.
$\phi \sim d v + 
const.$ which is regular on the future horizon, but singular on the past 
horizon. On the other hand $B=0$ and $d=A/2$ yields
 $\phi \sim u$ at infinity describing 
outgoing radiation. The same can be said for $\psi$.  Note that this
 behaviour emerges only as a  consequence of imposing a) and b) 
separately on $\langle  T_{\mu\nu}(\phi)\rangle$ and $\langle 
T_{\mu\nu}(\psi)\rangle$. Now, these auxiliary fields are related to the 
inverse of the fourth order operator $\Delta_4$ appearing in the nonlocal 
form of the action (\ref{14}). By our choice of constants we have 
therefore found the
boundary conditions appropriate to the description of 
black hole evaporation. \\
We come now to the numerology. As said before, $l_1$ is an arbitrary 
parameter of our model. If it is chosen such that $S^{eff}_{an}$ of eq. 
(\ref{14}) reduces to the Reigert action \cite{reigert} , i.e. 
$l_1=\frac{a}{8\pi\sqrt{-b}}$, inserting the appropriate values for one 
scalar field $(a=1/120 \ , b=-1/360 \ )$ we find 
$L=-\frac{1}{\pi(24M)^2 } $ which is negative. 
This is 
physically meaningless. This result is analogous to the one found for 
minimally coupled scalar fields classically reduced to 2D under spherical 
symmetry \cite{sfere}. On the other hand if $l_1=0$ which means, by our choice 
of constants, $\psi=0$ (i.e. the
conformally invariant part of $S^{eff}_{an}$ is completely removed)
 one 
gets $L= \frac{1}{720\pi M^2} $ 
which differs by a factor of $6 $ from the result 
\cite{dewitt}. The matching of this latter would require 
$\beta \sim\frac{5.8\ 10^{-1}}{\pi }$.  \\
Summarizing, we have shown that the characteristic behaviour at infinity 
and near the horizon of $\langle T_{\mu\nu}\rangle$ in the Unruh state 
for a Schwarzschild black hole on which our understanding of black hole 
evaporation so far is based can be reproduced by the anomaly 
induced effective action once appropriate boundary conditions are imposed 
on the auxiliary fields $\phi$ and $\psi$.  However, one should damp 
enthusiasm: $S^{eff}_{an}$ as it stands is not able to correctly 
reproduce subleading terms in $\langle T_{\mu\nu}\rangle$. For example 
one expects \cite{chrifu} that leading terms 
in $\langle T_{\theta}^{\ \theta}\rangle$ as $r\to\infty$
to start off as $r^{-4}$ 
whereas our analysis predicts the existence of a $r^{-3}$ term. This 
failure is not surprising given the incompleteness of $S^{eff}_{an}$. 
In particular, it is known that the Reigert action 
\cite{reigert} does 
not give the correct correlation functions of the theory \cite{eros}.
It would be interesting to consider some more complicated versions 
of the nonlocal effective action, which are based on the Green functions
of the second order conformal
operators rather than on the fourth order $\Delta_4$ .
\\
We end our work by mentioning that a similar construction 
can be given also
for the Hartle-Hawking state (black hole in thermal equilibrium) and 
for the Boulware state. We will report on this elsewhere \cite{bafash}.

R.B. wishes to deeply thank V. Frolov and R. Zucchini for discussions.
A.F. is supported by an INFN fellowship. I.Sh. is grateful to UFJF for
warm hospitality, to CNPq for the fellowship and to FAPEMIG (MG) 
for the travel grant. 
His work was partially supported by RFFI (project 99-02-16617).

\end{document}